# Real Differences between OT and CRDT in Correctness and Complexity for Consistency Maintenance in Co-Editors

DAVID SUN, Codox Inc., United States
CHENGZHENG SUN, Nanyang Technological University, Singapore
AGUSTINA NG, Nanyang Technological University, Singapore
WEIWEI CAI, Nanyang Technological University, Singapore

OT (Operational Transformation) was invented for supporting real-time co-editors in the late 1980s and has evolved to become core techniques widely used in today's working co-editors and adopted in industrial products. CRDT (Commutative Replicated Data Type) for co-editors was first proposed around 2006, under the name of WOOT (WithOut Operational Transformation). Follow-up CRDT variations are commonly labeled as "*post*-OT" techniques capable of making concurrent operations natively commutative in co-editors. On top of that, CRDT solutions have made broad claims of superiority over OT solutions, and often portrayed OT as an incorrect and inefficient technique. Over one decade later, however, CRDT is rarely found in working co-editors; OT remains the choice for building the vast majority of today's co-editors. Contradictions between the reality and CRDT's purported advantages have been the source of much confusion and debate in co-editing researcher and developer communities. To seek truth from facts, we set out to conduct a comprehensive and critical review on representative OT and CRDT solutions and working co-editors based on them. From this work, we have made important discoveries about OT and CRDT, and revealed facts and evidences that refute CRDT claims over OT on all accounts. These discoveries help explain the underlying reasons for the choice between OT and CRDT in the real world. We report these results in a series of three articles.

In this article (the second in the series), we reveal the differences between OT and CRDT in their basic approaches to realizing the same general transformation and how such differences had resulted in different technical challenges and consequential correctness and complexity issues. Moreover, we reveal hidden complexity and algorithmic flaws with representative CRDT solutions, and discuss common myths and facts related to correctness and complexity of OT and CRDT. We hope the discoveries from this work help clear up common myths and confusions surrounding OT and CRDT, and accelerate progress in co-editing technology for real world applications.

KEYWORDS

Operational Transformation (OT); Commutative Replicated Data Type (CRDT); Concurrency Control; Consistency Maintenance; Real-Time Co-Editing; Cloud/Internet/Distributed Computing; Computer Supported Cooperative Work (CSCW) and Social Computing.

## 1   INTRODUCTION

Real-time co-editors allow multiple geographically dispersed people to edit shared documents at the same time and see each other's updates instantly [1,7,15,17,18,19,41,46,59,60,65,77,83]. One major challenge in building such systems is consistency maintenance of documents in the face of concurrent editing, under high communication latency environments like the Internet, and without imposing interaction restrictions on human users [15,59,60].

Operational Transformation (OT) was invented to address this challenge in the late 1980s [15, 59,66,77]. OT introduced a framework of transformation-based solutions to ensure consistency in the presence of concurrent user activities. The OT framework is grounded in established distributed computing theories and concepts, principally in *concurrency* and *context* theories [27, 59,71,72,88]. Since its inception, the scope of OT research has evolved from the initial focus on consistency maintenance to include a range of key collaboration-enabling capabilities, including *group undo* [41,47,62,63,71,72], *workspace awareness* [1,22,65], etc. In the past decade, a main impetus to OT research has been to move beyond plain-text co-editing [7,15,23,41,46,59,60,63,66, 75,76,82], and to support real-time collaboration in rich-text co-editing in word processors [65,70, 73,87], Web document co-editing [12], spreadsheet co-editing [74], 3D model co-editing in digital media tools [1,2], and file synchronization in cloud storage systems [3]. OT-based co-editors have also evolved from supporting people to use the same editor in one session (*homogeneous* co-editing), to allowing the use of different editors in the same session (*heterogeneous* co-editing) [10]. Recent years have seen OT being widely adopted in industry products as the core technique for consistency maintenance, ranging from battle-tested online collaborative rich-text editors like Google Docs[1][13], to emerging start-up products, such as Codox Apps[2].

A variety of alternative techniques for consistency maintenance in co-editors had also been explored in the past decades [17,19,21,44,45,77]. One notable class of techniques is CRDT[3] (Commutative Replicated Data Type) for co-editors [4,5,9,28,35,40,42,43,44,48,50,51,84,85,86]. The first CRDT solution for plain-text co-editing appeared around 2006 [43,44], under the name of WOOT (WithOut Operational Transformation). One motivation behind WOOT was to solve the *FT (False Tie)* puzzle in OT [58,60] (discussed in Section 3.3 and Section 5), using a radically different approach from OT. Since then, numerous WOOT revisions (e.g. WOOTO [85], WOOTH [4]) and CRDT alternatives (e.g. RGA [48], Logoot [84,86], LogootSplit [5]) have appeared in literature, mostly confined in the domain of plain-text co-editing. In CRDT literature, CRDT has commonly been labelled as a "*post*-OT" technique that makes concurrent operations *natively* commutative, and does the job "*without operational transformation*" [43,44], and even "*without concurrency control* " [28]. On top of that, CRDT solutions have made broad claims of superiority over OT solutions, and routinely portrayed OT as an incorrect and inferior technique[4].

After over one decade, however, CRDT solutions are rarely found in working co-editors or industry co-editing products, and OT solutions remain the choice for building the vast majority of co-editors. The contradictions between the reality and CRDT's purported advantages over OT have been the source of much confusion and debate in co-editing research and developer communities (see footnote 4). Have the majority of co-editors been unfortunate in choosing the faulty and inferior OT, or such CRDT claims are false? What are the real differences between OT and CRDT for co-editors? What are the key factors that may have affected the choice between OT and CRDT for co-editors in the real world? We believe that a thorough examination of these questions is relevant not only to researchers exploring the frontiers of collaboration-enabling technologies and systems, but also to practitioners seeking viable techniques to build real world collaboration tools and applications.

---

[1] https://www.google.com/docs/about/

[2] https://www.codox.io

[3] In literature, CRDT can refer to a number of different data types [50,51]. In this paper, we focus exclusively on CRDT solutions for text co-editors, which we abbreviate as "CRDT" in the rest of the paper, though occasionally we use "CRDT for co-editors" for emphasizing this point and avoiding misinterpretation.

[4] We posted an early version of our report on this work at https://arxiv.org/abs/1810.02137, in Octo. 2018, which attracted wide interests and discussions in public blogs (among academics and practitioners) and private communications (between readers and authors). The following link, at https://news.ycombinator.com/item?id=18191867, hosts some representative comments and opinions on various co-editing issues addressed in our article. One well-known CRDT advocate commented there: *"The argument of Sun's paper seems to be that CRDTs have hidden performance costs. Perhaps this is true. This completely misses the main point. OT is complex, the theory is weak, and most OT algorithms have been proven incorrect (…). AFAIK, the only OT algorithm proved correct is TTF, which is actually a CRDT in disguise. In contrast, the logic of CRDTs is simple and obvious. We know exactly why CRDTs converge. … Disclaimer: I did not read the paper in detail, just skimmed over it."* This CRDT advocate basically re-iterated some common CRDT claims against OT, which re-confirms the liveness of ongoing debates, and warrants a thorough examination of those CRDT claims. Without reading the paper in detail, this CRDT advocate clearly missed the facts and arguments presented in our paper. In fact, we had examined all points mentioned above (and beyond), and revealed facts and evidences that refute those CRDT claims on all accounts. Readers may make independent judgement after reading our papers in this series.

To seek truth from facts, we set out to conduct a comprehensive and critical review of representative OT and CRDT solutions, and working co-editors based on them, which are available in publications or from publicly accessible open-source project repositories. In this work, we explored *what*, *how*, and *why* OT and CRDT solutions are different and the consequences of their differences from both an algorithmic angle and a system perspective. From this exploration, we made a number of discoveries, some of which are rather surprising. One such discovery is that CRDT is not natively commutative for concurrent operations for co-editors as commonly claimed, but is the same as OT in following a general transformation approach in co-editors, albeit indirectly. This study has also examined major CRDT claims over OT and revealed facts and evidences that refute those claims on all accounts.

We have focused this study on OT and CRDT solutions to *consistency maintenance* in *real-time* co-editing, as it is the foundation for other co-editing capabilities, like group *undo* and issues related to *non-real-time* co-editing, which we plan to cover in future work. Also, we focus our discussions on basic issues in plain-text co-editing as the capability of supporting plain-text co-editing is the foundation for support co-editing in more advanced domains (e.g. rich text co-editing), and the vast majority of published CRDTs are confined in this domain. We know of no existing work that has made similar attempts.

The topics and bulk of outcomes from this study are comprehensive, complex and diverse, and have different accessibilities to readers with different interests and backgrounds. To cope with the complexity and diversity, and take into account of feedback to a prior version of our report on this work (see footnote 4), we have organized the outcome materials into three parts and presented them in a series of three related but self-contained articles.

The current paper is the second in this series. We focus on examining the real differences between OT and CRDT in correctness and complexity for consistency maintenance in co-editors. We dissect representative OT and CRDT solutions, and explore how different basic approaches to realizing the same transformation – the *concurrency-centric* and *direct* transformation approach taken by OT versus the *content-centric* and *indirect* transformation approach taken by CRDT – had resulted in different technical challenges, and consequential correctness and complexity issues. We also reveal hidden algorithmic flaws with representative CRDTs, and discuss common myths and facts related to correctness, complexity, and simplicity of OT and CRDT solutions.

The rest of this paper is organized as follows. In Section 2, we briefly review the general transformation (GT) framework to consistency maintenance in co-editors, and the basic approaches taken by OT and CRDT in realizing the same GT, which provides the necessary background results reported in the prior paper [78]. In Sections 3 and 4, we examine the key technical issues and solutions in OT and CRDT, respectively, rooted in their basic approaches to realizing the GT. In Section 5, we discuss and clear up common myths and misconceptions associated with OT and CRDT in correctness and complexity. In Section 6, we conclude this paper by summarizing the main results reported in this paper.

## 2    GENERAL TRANSFORMATION AND DIFFERENT OT AND CRDT REALIZATIONS

In [78], we have presented a general transformation (GT) framework for consistency maintenance in co-editors, and revealed that OT and CRDT are different realizations of the same GT. Moreover, we found that CRDT is not natively commutative for concurrent operations in co-editors, as often claimed (a myth) in CRDT literature. Uncovering the hidden transformation nature and demystifying the commutativity property of CRDT provides the foundation to reveal real differences between OT and CRDT for co-editors in correctness and complexity, which is the focus of this paper. For background information, we briefly review the basic ideas of the GT and illustrate OT and CRDT different realizations of the same GT in this section.

## 2.1 Basic Ideas of the General Transformation

In Fig. 1-(a), we illustrate the basic ideas of the GT under a text co-editing scenario, with two co-editing users A and B. We summarize some key GT characteristics below:

1. **Replicated document states:** The shared document is replicated at all co-editing sites. In this example, a string "abe" is replicated at both sites.
2. **Unconstrained interaction:** Users are allowed to freely and concurrently edit any parts of the local document without blocking, locking or other restrictions. In this example, two operations are generated: $O_1 = D(1)$ (to delete the character at position 1) generated by User A; $O_2 = I(2, "c")$ (to insert a character "c" at position 2) generated by User B.
3. **Fast local response:** Local operations are executed as-is immediately. In this example, the two local operations are executed on the local documents without any delay.
4. **Real-time notification:** Local operations are propagated to remote sites quickly so that users can see each other's edits lively. Notification latency is constrained by external network communication delay only, without any GT synchronization cost.
5. **Consistency maintenance by transformation:** Remote operations are transformed before execution for consistency maintenance. The transformation may change parameters of a remote operation according to the effects of concurrent operations (if any). In this example, when $O_2$ arrives at User A, it is transformed (by any GT solution) into $O_2' = I(1, "c")$, where the original position parameter 2 in $O_2$ has been changed to 1 in $O_2'$, to compensate the position-shifting effect of the previously executed concurrent operation $O_1$. If $O_2$ were executed without such transformation, incorrect editing results would have been produced due to position-shifting effects among concurrent operations, which is the problem the GT is to solve. When $O_1$ arrives at User B, it is transformed as well, but the resulting $O_1'$ is exactly the same as the original $O_1$ as the previously executed $O_2$ has no shifting effect on $O_1$. The underlying GT solution should be able to determine whether and how to change the parameters of an operation.
6. **Convergence and intention preservation as consistency requirements:** After executing the same group of concurrent operations at all co-editing sites, the document states at all sites are consistent in the sense of being *convergence* (replicas are identical) and *intention-preserving* (the original effect of an operation on the local replica are preserved at all remote replicas) [60]. In this example, after executing $O_1$ and $O_2$ at the two sites, the replica states converge into the same "ace", and this identical state also preserves the editing effects (or *intentions*) of both $O_1$ and $O_2$.
7. **Achieving commutativity**: Concurrent operations are made *commutative*, meaning they can be executed in different orders, but achieve the same results. For example, $O_1$ and $O_2$ are executed in different orders and the same state "ace" is achieved.

The co-editing scenario illustrated in Fig. 1-(a), and the sketched GT characteristics should be familiar to readers with some background in OT. Indeed, this scenario has often been used to explain basic ideas of OT, and these GT characteristics were originated from OT literature as well [15,59,60]. What might be surprising to many is that the same co-editing scenario, inconsistency issue involved, and basic GT ideas apply equally to CRDT as well.

## 2.2 Different Realizations of the General Transformation

We instantiate the scenario under the GT approach (Fig. 1-(a)) with OT and WOOT (the first and representative CRDT) realizations in Fig. 1-(b) and (c), respectively. We use *External State* (ES) to denote replicated document states visible to users, and *External Operation* (EO) to represent operations generated by users. In fact, ES and EO are exactly the same as the document states and operations illustrated in Fig.1-(a) for GT. Clearly, OT and WOOT share the same ES and EO, as well as the same inconsistency issue caused by position-shifting effects of concurrent operations.

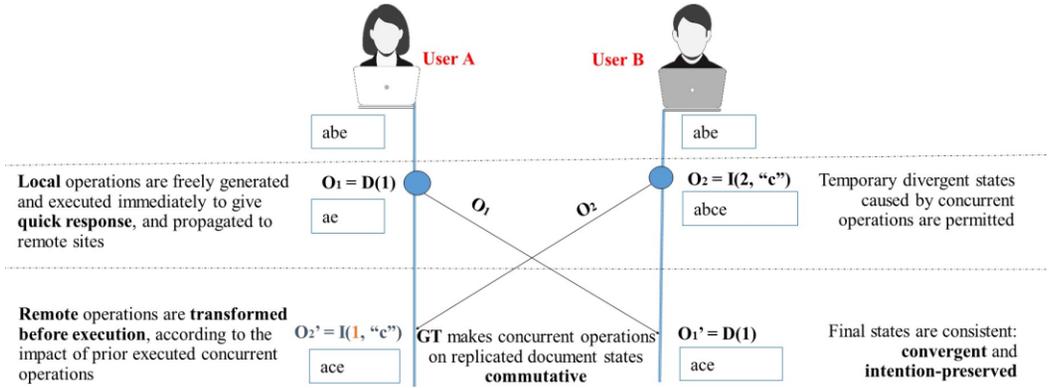

(a) Basic idea of the general transformation (GT) for consistency maintenance in co-editors.

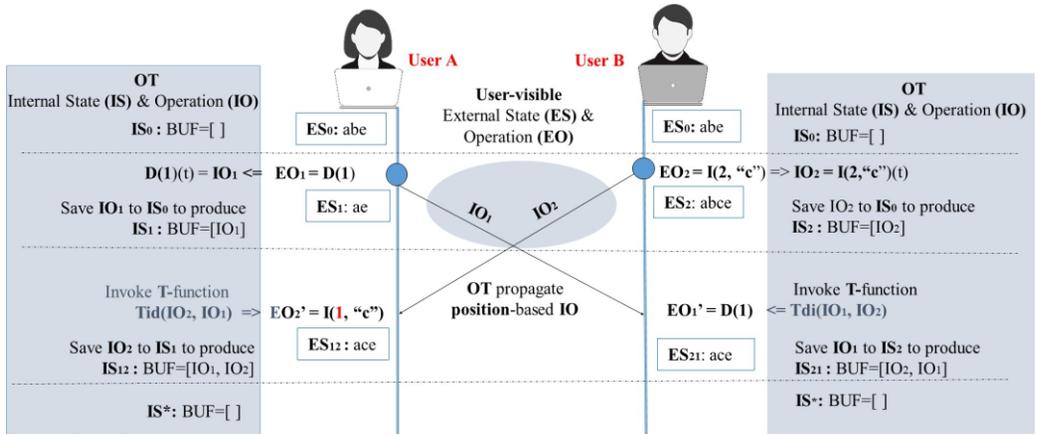

(b) The OT approach to realizing the general transformation.

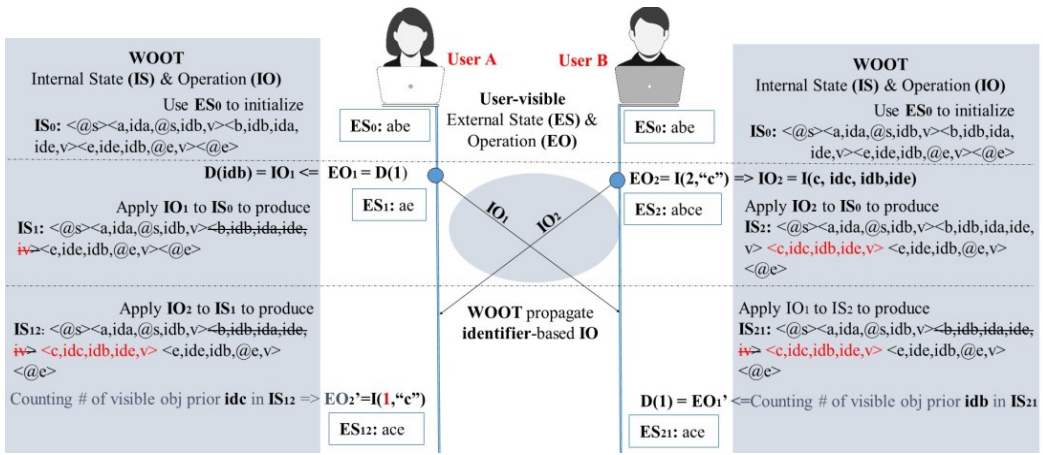

(c) The WOOT (CRDT) approach to realizing the general transformation.
Note: <@s> and <@e> are two special objects marking the *start* and *end* points of an internal state.

Fig. 1. General transformation and different realizations by OT and WOOT (CRDT).

For consistency maintenance, OT and WOOT have devised different internal data structures and mechanisms, in the form of *Internal State* (IS) and *Internal Operation* (IO), which are highlighted in shaded regions of Fig. 1-(b) and (c). It should be stressed that IS and IO are used internally only, and invisible to editors or users.

In general, IS serves the purpose of keeping track of the concurrency impact information needed for realizing the GT; IO presents the effect of the corresponding EO, and captures concurrency relationships among operations. For OT, IO is *a timestamped EO*, e.g. $D(1)(t)$ and $I(2,"c")(t)$ in Fig. 1-(b) (note: IO and EO are all *position*-based operations in OT); IS is a buffer of internal operations to record concurrency impact information. In contrast, WOOT IS contains a sequence of objects, which mirrors the sequence of chars in ES, plus tombstones for representing deleted chars; WOOT IO is based on specially designed *immutable identifiers*, e.g. $D(idb)$ and $I(c, idc, idb, ide)$ in Fig. 1-(c), which get rid of the positional parameter of the corresponding EO. For WOOT to work, an insert operation carries not only the identifier of the target object (i.e. the new character to be inserted), but also identifiers of two neighboring objects corresponding to the characters in ES that are visible to the user at the time when the insert was generated. The target identifier and neighboring object identifiers, together with tombstones embedded in the object sequence, are crucial elements in WOOT's solution to concurrency issues and the FT puzzle [58]. In Section 4, we give more elaborations on various CRDT identifiers.

Based on respective IS and IO, OT and WOOT have devised different ways to handle local and remote operations. The general workflow of handling an operation from local generation to remote replay is sketched in Fig.1-(b) and (c). Briefly, we highlight some key observations: both OT and WOOT produce the same final position-based operations, i.e. $EO_2 = I(2, "c")$ is transformed into $EO_2' = I(1, "c")$ at User A, while $EO_1 = D(1)$ is unchanged at User B. For detailed explanations of the internal workings of OT and WOOT, we refer readers to Section 3.1.2 and Section 3.2.2, respectively, in [78]. This intuitive example shows WOOT is indeed a realization of the GT.

In general, all CRDT solutions have followed the same GT approach, albeit indirectly: CRDT first converts a position-based EO into an identifier-based IO at the local site, and then convert the identifier-based IO back to the position-based EO at the remote site. In contrast, OT realizes the GT directly by transforming position-based operations into position-based operations.

In summary (see Table 1), both OT and CRDT aim to address the same issues and achieve the same consistency defined as convergence and intention-preservation, share the same ES and EO for text editing, follow the same GT approach, and make the same concurrent EOs commutative by transformation, not natively. However, OT and CRDT have devised different IO and IS, and taken different (*direct* vs *indirect*) approaches to realizing the GT, as highlighted in shaded rows. The different approaches taken by OT and CRDT in realizing the GT had resulted in different technical challenges and had major impacts on the correctness, complexity and efficiency of OT and CRDT solutions, which are revealed in detail in the rest of this paper.

Table 1 High-level similarities and differences (in shaded rows) between OT and CRDT in realizing GT

|  | OT | CRDT |
|---|---|---|
| Consistency Requirements | Convergence and intention-preservation ||
| External State (ES) | A sequence of characters: $c_0, c_1, c_2, ..., c_n$ ||
| External Operation (EO) | A pair of position-based operations: *Insert(p, c)* or *Delete(p)* ||
| General Transform (GT) | GT: $EO_{local} \rightarrow EO_{remote}$ ||
| Commutativity | Concurrent EOs are made commutative by GT, not natively ||
| Internal Operation (IO) | Position-based operations by timestamping EO | Immutable identifier-based operations from converting EO |
| Internal State (IS) | A buffer of concurrent IOs, independent of ES | A sequence of objects (+ optional tombstones), mirroring chars in ES |
| GT Realization | Direct transformation: position $\rightarrow$ position | Indirect transformation: position $\rightarrow$ identifier $\rightarrow$ position |

## 3  KEY TECHNICAL ISSUES AND SOLUTIONS WITH OT

Technically, OT solutions[5] are concurrency-centric in the sense they treat generic concurrency issues among operations with the first priority at the core control algorithms, and separately handle application-specific data and operation issues within transformation functions. One fundamental impact of this concurrency-centric approach is, among others, OT time and space complexities are determined by a variable c (for concurrency), which represents the number of concurrent operations involved in transforming an operation.

### 3.1  Control Algorithms

*3.1.1 The dOPT Puzzle and the Theory of Operation Context.* Designing correct and efficient OT control algorithms was a major challenge in early OT research [59]. Under the first control algorithm – dOPT (distributed OPerational Transformation) [15], two operations could be transformed with each other as long as they had a concurrency relationship, which turned out to be inadequate. This algorithmic flaw, named the dOPT *puzzle* later, was subtle and had taken a few years for several researchers to independently discover and resolve [59].

The key to resolving this dOPT puzzle is to ensure the two input operations to a transformation function are not only concurrent, but also defined on the same document state, or an *equivalent context* [60]. With the guarantee of *context-equivalence,* a transformation function can compare parameters of input operations to derive their concurrency-impact on each other. Detecting and resolving the dOPT puzzle has led to the invention of multiple OT control algorithms capable of ensuring the context-equivalence condition, and the establishment of the theory of *operation context* [59,71,72], which become a cornerstone of OT correctness. For a comprehensive review of independent solutions to the dOPT puzzle, the reader is referred to [59].

*3.1.2 Vector versus Scalar-based Timestamps.* To capture concurrency and context relationships among operations, an OT solution may use *vectors* (with one element for each of co-editing sites in a session) or *scalars* (with a fixed number of variables) for timestamping operations or control purposes. For example, some OT solutions, including adOPTed [46], GOT [58,60], GOTO [59], SOCT2 [56], and COT [71,72], have used vector-based timestamps; but other OT solutions, such as Jupiter [39], NICE [52], TIBOT [29,89], SOCT4 [82], Google Wave and Docs OT [13,37,83], and POT [89], have used scalar-based timestamps.

*3.1.3 Server-based versus Distributed OT.* Numerous OT control algorithms have been designed with a variety of system architectures and communication topologies. Some OT control algorithms, such as Jupiter, NICE, and Google OT, are *Severed*-based OT (SOT) algorithms as they require a central transformation server and co-editing clients must communicate through the server; but most OT control algorithms, from the first dOPT, to adOPTed, GOT, GOTO, SOCT2, COT, TIBOT, and POT, are *Distributed* OT (DOT) algorithms, which require no transformation server and allow co-editing clients to be connected with each other in any suitable communication topologies (with or without a server). In co-editing literature, one common misconception is to label OT as a technique that always requires a server, which led to the misconception that OT is unsuitable for *peer-to-peer* co-editing (see more discussion on this point in [78]).

*3.1.4 Complexity in Theory.* Some OT control algorithms, including adOPTed, GOT, GOTO, and SOCT2, had the quadratic time complexity in transforming a remote operation – $O(c^2)$, where $c$ is the number of concurrent operations involved in transformation. Though this theoretic complexity was not a practical concern in real-time co-editors (see Section 3.1.5), more efficient control algorithms were proposed, e.g. Jupiter, NICE, TIBOT, SOCT4, Google Wave and Docs OT, COT and POT, which have a time complexity of $O(c)$ for transforming a remote operation. For processing local operations[6], OT solutions generally have the constant time complexity $O(1)$.

---

[5] In this paper, we focus *exclusively* on OT solutions that separate *generic* control algorithms from *application-specific* transformation functions [1-3,7,12,13,15,18,23,29,34,37,39,41,46,47,52-63,65,70-77,81-83,87-89], as they represent the majority and mainstream OT solutions, on which existing OT-based co-editors are built. In literature, however, there are other OT solutions (e.g. [30-33,49]), in which control procedures are not generic but dependent on specific types of operation and data, and transformation functions may examine operation concurrency relationships as well. In those OT solutions, "*control procedure and transformation functions are not separated as in previous works – instead, they work synergistically in ensuring correctness*" [33], and different correctness criteria were used as well [30-33,49,66].

For recording concurrency-impact information, an OT solution keeps a buffer of operations. The space complexity of this buffer is characterized by two variables $c$ and $m$, where $c$ is the same as above, and $m$ is the number of users in a session, and depends on whether using scalars or vectors for timestamping operations, and whether maintaining single or multiple transformation paths in the buffer, as summarized in Table 2.

Table 2 Space complexities in representative OT solutions

|  | Single-Transformation-Path | | Multi-Transformation-Path |
|---|---|---|---|
| **Scalar Timestamps** | $O(c)$: | Google OT [13,83], Jupiter[39], NICE[52], TIBOT[29], SOCT4[82] | $O(c \cdot m)$: POT [89] |
| **Vector Timestamps** | $O(c \cdot m)$: GOT[58,60], GOTO[59], SOCT2[56] | | $O(c \cdot m^2)$: adOPTed [46], COT[72] |

*3.1.5 Complexity in Practice.* Due to its concurrency nature, the variable $c$ has two properties: (1) $c$ is often bounded by a small value, e.g. $c \leq 10$, in real-time co-editing sessions, in which the number ($m$) of users is small (e.g. $m \leq 5$), and the number of operations each user may generate per second is small (e.g. $\leq 2$) due to the relative slow pace of human interactions and operation composing schemes commonly used in real-time co-editors [66,83]; (2) $c$ can be reduced to *zero* by garbage collection schemes that remove buffered operations whenever it is no longer possible for them to be concurrent with future operations [60,66,72,89]. Garbage collecting non-concurrent operations has been well-established not only in theory, but also commonly adopted in OT-based co-editors, including Google Wave and Docs [13,83], CoWord [65], Codox Apps [78], etc.

## 3.2 Transformation Functions

With control algorithms at the core, an OT solution has transformation functions as the application-specific part. For correctness of an OT solution, transformation functions are required to meet certain transformation properties as well. Establishing general transformation properties for correctness is one major achievement in past OT research [15,41,46,59,60,66,71,72,75,76,89].

*3.2.1 Convergence Properties CP1 and CP2.* Two transformation properties are directly relevant to convergence preservation.

**Convergence Property 1 (CP1):** Given $O_a$ and $O_b$ defined on the document state *DS*, and a transformation function *T*, if $O_a' = T(O_a, O_b)$, and $O_b' = T(O_b, O_a)$, the following holds:
$$DS \circ O_a \circ O_b' = DS \circ O_b \circ O_a',$$
which means applying $O_a$ and $O_b'$ in sequence on *DS* produces the same state as applying $O_b$ and $O_a'$ in sequence on *DS*.

Preserving CP1 is a key for an OT solution to make concurrent editing operations commutative. Numerous transformation functions capable of preserving CP1 (under plain-text and other document models) have been designed [1,2,3,46,74,75,76].

---

[6] The local operation processing time covers the period during which the local operation is timestamped and saved in the buffer (ready for propagation), as sketched in OT Local Operation Handler (LOH) under the general transformation framework in Table 1 of [78]. For most OT solutions, a local operation in the buffer is propagated as-is without further processing. For a few OT solutions, notably Google Wave and Docs OT, TIBOT, and SOCT4, local operations may wait in the buffer until certain conditions are met, e.g., a local operation is not propagated until the prior propagated local operation has been acknowledged (for Google Wave and Docs OT); a local operation is not propagated until all remote operations with *time-interval-based* timestamps (for TIBOT) or *global sequence numbers* (for SOCT4) that are earlier than that of the waiting local operation have been received and processed. While waiting in the buffer, a local operation may be transformed with incoming remote operations, and such processing time is part of handling a remote operation by Remove Operation Handler (ROH) in Table 1 of [78] (with the complexity $O(c)$). Handling a local operation is completed at the moment when an OT solution is able to handle another local operation (by LOH) *or* remote operation (by ROH).

Further research found that CP1 alone may not be sufficient to ensure convergence (for OT-based co-editors supporting more than 2 users); an additional property CP2 may be required under certain conditions (more elaboration on those conditions in Section 3.3 and Section 5) [46,72,89].

**Convergence Property 2 (CP2):** Given $O_a$, $O_b$ and $O_c$ defined on the same state, and a transformation function $T$, if $O_c' = T(O_c, O_b)$ and $O_b' = T(O_b, O_c)$, the following holds:
$$T(T(O_a, O_b), O_c') = T(T(O_a, O_c), O_b'),$$
which means transforming $O_a$ against $O_b$ and then $O_c'$ produces the same operation as transforming $O_a$ against $O_c$ and then $O_b'$.

It is worth highlighting that the transformation function $T$, document state $DS$, and operation $O$ are all unspecified in CP1 and CP2 specifications, meaning CP1 and CP2 are generally applicable to transformation functions defined for any document states and operation models.

*3.2.2 The CP2 Issue and the False-Tie (FT) Puzzle.* Unlike CP1, which is relatively easy to preserve in plain-text co-editing, CP2 is non-trivial to preserve under certain circumstances. One CP2-violation case in plain-text co-editing, named as the FT puzzle, was first detected and reported in [58,60]. As a matter of fact, the FT puzzle was not reported as a major result in [58,60], which were mainly about solving the dOPT puzzle, but presented as a counter-example to show that CP2 could be violated by the specific transformation functions proposed in the context of the adOPTed algorithm in [46].

It turned out that the discovery of the FT puzzle had enormous impact on follow-up development of OT, the invention of WOOT and follow-up CRDTs, and other OT alternatives as well (more discussions in Sections 4 and 5) [24,43,44,45].

### 3.3 Solving the CP2 Issue and the FT Puzzle under the OT Approach

The CP2-violation issue has been solved under the OT approach in various ways, which are elaborated in this section.

At the start, we highlight one fundamental point: CP2 is not unconditionally required, but required for transformation functions *only if* the control algorithms in an OT solution select concurrent operations for transformation in arbitrary orders (e.g. the adOPTed algorithm in [46]), or more precisely, transform the same pair of concurrent operations under different contexts [46,66,71,72,89]. This precondition for CP2 has played a key role in the invention and verification of OT solutions to the CP2 issue along two alternative paths.

First, the CP2 issue can be addressed by designing OT control algorithms capable of avoiding transforming operations in arbitrary orders – the CP2-avoidance approach, which is generic and independent of operation types and data models. When control algorithms with the CP2-avoidance capability are used in an OT solution, CP2 is no longer a correctness requirement for corresponding transformation functions. Control algorithms capable of avoiding CP2 are numerous [13,29,37,39,52,55,71,72,82,83,89].

It should be stressed that CP2-avoidance solutions impose no restriction on the user's ability to freely and concurrently edit the shared document, and are designed with or without using a server for operation transformation and/or propagation, and using either vectors or scalars for timestamping operations. It is mistaken to label CP2-avoidance solutions as being based on servers or unsuitable for peer-to-peer co-editing (as often found in CRDT publications, e.g. [4]). A comprehensive study of underlying conditions and patterns of seven different CP2-avoidance control algorithms is available in [88,89].

Second, the CP2 (and FT) issue can also be addressed by designing transformation functions capable of preserving CP2, without changing underlying operation and data models (which is in contrast to TTF [45], to be discussed in Section 5.2). This CP2-preserving approach is suitable to OT solutions with control algorithms that transform operations in arbitrary orders (e.g. adOPTed [46], GOTO [59], and SOCT2 [56]). Obviously, transformation functions capable of preserving CP2 (and CP1) have no problem to be used in combination of control algorithms with the CP2-avoidance capability. Publically available and verified transformation functions capable of preserving CP2 (and CP1) for string-wise data and operation models can be found in [76].

Among these two alternative approaches to addressing CP2, the CP2-avoidance approach is often favored as CP2-avoiding control algorithms are generally applicable to data and operation models for a wide range of co-editing applications beyond plain-text editing. OT solutions for advanced co-editing applications, such as rich-text word processors [65,70,87], 2D spreadsheets [74], 3D digital media design systems [1,2], and shared workspaces in cloud storage systems [3], have mostly used a combination of transformation functions for preserving CP1 (plus application-specific combined effects for concurrent operations [1,3,66,75]) and control algorithms for avoiding CP2[72,88,89].

Apart from above CP2-avoidance or CP2-preserving strategies, there were other attempts to address the FT puzzle and the CP2-violation issue [24,25,30,31,32,33,43,44,45,57]. We discuss them, together with misconceptions surrounding FT and CP2, in Section 5.

### 3.4  Summary of Correctness and Complexity of OT Solutions

In over two decades, numerous OT solutions have been invented to address OT-special technical challenges, including the dOPT puzzle and the FT puzzle, and most importantly, to support building real world co-editors from text editing to more advanced and complex editing applications. The correctness of key OT components, including generic control algorithms and transformation functions for a range of commonly used operation and data models (e.g. string-wise plain-text editing and beyond) [1,3,74,75,76], has been established under well-defined conditions and properties, including CP1 and CP2 [15,34,41,46,59,60,63,66,72,76,89]. State-of-the-art OT solutions have achieved the space complexity $O(c)$ or $O(c \cdot m)$, and the time complexity $O(1)$ and $O(c)$ for processing local and remote operations, respectively.

## 4   KEY TECHNICAL ISSUES AND SOLUTIONS WITH CRDT

In over one decade, CRDT has evolved from the initial WOOT to a myriad of CRDT solutions, mostly confined in plain-text co-editing [4,5,9,28,35,40,42,43,44,48,50,51,84,85,86]. CRDT solutions have been commonly designed around internal object sequences, object identifiers, identifier-based operations, and schemes for searching and manipulating internal object sequences. This CRDT design approach is *content-centric* in the sense that it focuses on manipulating *contents*, such as an object sequence (with or without tombstones) and identifiers, which are directly related to application data and operation models. In effect, CRDT solutions treat application-specific data and operation issues with the first priority, but mix concurrency issues within identifier creation, object search and manipulation schemes.

At first glance, the CRDT object sequence may appear simple and easy to understand as anyone with basic knowledge of data structures and algorithms in computer science would know how to manipulate linear object sequences, without necessarily knowing advanced techniques in distributed computing and co-editing. Unfortunately, this is an illusion. While linear object sequences and search algorithms in sequential processes are well-understood in computer science, understanding such data structures and algorithms under real-time co-editing scenarios is a different matter altogether. It turned out that understanding various issues in real-time co-editing is a big challenge – a lesson learnt from past co-editing research history [59], and repeatedly confirmed by numerous CRDT proposals, which are examined in this section.

Fundamentally, concurrency issues are inherent in unconstrained co-editing, and common to both OT and CRDT, as illustrated in Section 2. CRDT solutions have to face and handle concurrency issues in content-specific ways, meaning to mix concurrency solutions within specific object sequences, identifiers, and search and manipulation schemes, leading to CRDT-special correctness and complexity problems.

### 4.1  $C/C_t$-based Complexities in CRDT Solutions

A direct consequence of the CRDT content-centric approach is that time and space complexities of CRDT solutions are inherently determined by a variable $C$ (for *Contents*) or $C_t$ (for *Contents* *with tombstones*) – the number of objects in the internal object sequence, which mirrors the external sequence of characters in the user-visible document. *C* often takes a big value, e.g. $10^3 \leq C \leq 10^6$, for common text document sizes ranging from 1KB to 1MB; $C_t$ is much bigger than *C* due to tombstones (more elaboration on this in Section 4.2). The CRDT object sequence (as a whole, not just tombstones) is CRDT-special space overhead; it is incorrect to portray the object sequence overhead as generally required for text co-editing in some CRDT papers (e.g. [6]).

In Table 3, we give a summary of objects, identifiers, time and space complexity, and other characteristic features of some representative CRDT solutions, which are elaborated below.

*4.1.1 WOOT with Time Complexity $O(C_t^3)$.* In WOOT, to achieve object sequence consistency in the face of concurrent inserts, designers had to devise a recursive procedure *IntegrateIns* to search the object sequence in multiple nested loops in order to determine the correct insertion location. The *IntegrateIns* procedure was quite intricate to devise and had taken the designers multiple iterations to get it right, as reported in [43], which resulted in an algorithm with the time complexity of $O(C_t^3)$ [43,44]. Understanding how the *IntegrateIns* procedure works and verifying its correctness (still missing in CRDT literature) require intimate knowledge of real-time concurrent co-editing, which is far beyond just knowing sequential object search algorithms. Indeed, a later WOOT-like CRDT solution [40] got it wrong when trying to alter WOOT's ways of handling neighboring objects and tombstones (more elaboration in Section 4.2).

*4.1.2 WOOTO and WOOTH with Time Complexity $O(C_t^2)$.* The WOOT $O(C_t^3)$ complexity was obviously too costly for executing a single insert operation (at both the local and remote object sequences), so various improvements were proposed, including WOOTO[85], which used a *degree* scheme to capture the relative ordering of concurrent object creations and save one round of object sequence search; and WOOTH [4], which used a *hash* scheme to speed up the search of neighboring objects. Both WOOTO and WOOTH achieved the time complexity $O(C_t^2)$, which is better than $O(C_t^3)$, but still far from being cheap given the big value of $C_t$.

*4.1.3 RGA with Time Complexity $O(C_t)$ or $O(C)$.* RGA (Replicated Growable Array) was yet another CRDT solution for co-editing [48], which also adopted the tombstone-based approach to representing the internal object sequence, but used a hash scheme and vector-based total and causal ordering to speed up the search of the target object or location in the object sequence. RGA achieved the time complexity of $O(C_t)$ (or $O(C)$ with tombstone garbage collection, to be discussed below) for executing an insert operation, but with additional costs in maintaining the hash table and issues related to vector clocks [43,44,84]. RGA is often quoted as the fastest tombstone-based CRDT solution (e.g. [4]).

*4.1.4 Object Sequence Search with Time Complexity $O(C)$ or $O(log(C))$.* The cost of applying an identifier-based operation in the object sequence (e.g. the cost of *IntegrateIns*) is only one part of handling a local or remote operation in WOOT (and its variations). Another often-occurring cost is searching the object sequence, which has the time complexity $O(C_t)$, and may occur at a number of places, e.g. converting a position-based operation into an identifier-based operation at the local site (using the position as the search key), checking the existence of an object in the sequence for determining whether to accept a remote operation (using the identifier as the key), and converting an identifier-based operation to a position-based operation for replay at a remote site (counting visible objects in the sequence, as illustrated in Fig. 1-(c)), etc.

A later paper [9] tried to address the CRDT *poor responsiveness*[7] in the *upstream* processing (i.e. processing local operations). This paper attributed the problem to the time complexity $O(C_t)$ or $O(C)$ for the search process in converting a position-based operation into an identifier-based operation. The proposed solution was to add one extra binary tree to speed up the local search.

---

[7] In [9], the authors acknowledged "*upstream execution – and thus responsiveness of CRDT algorithms often performs poorly,*" but claimed "*downstream execution of CRDT algorithms is more efficient by a factor between 25 to 1000 compared to representative OT algorithms*", quoting the evaluation results in [4], in which the referred representative OT algorithm is the TTF solution [45], which is "*actually a CRDT in disguise*" (see footnote 4). We examine TTF in Section 5.2.

Table 3. Key features of representative CRDT solutions. $C_t$ and $C$ represent the number of objects in the internal object sequence with or without tombstones, respectively. Time complexity is for handling one *insert* operation internally, which incurs at both local and remote sites. G.C. stands for Garbage Collection of tombstones.

| | WOOT [43,44] | WOOTO [85] | WOOTH [4] | RGA [48] | Logoot [84] | Logoot-Undo [86] |
|---|---|---|---|---|---|---|
| **Object sequence** | @s, o₀, o₁, ..., oₙ, @e where $o_x$ is an object corresponding to a character in external document or a tombstone; @s and @e are special objects marking the start and end of the object sequence. | | | | | |
| | **tombstones** kept in the object sequence | | | | **no tombstone** kept | |
| **Identifier-based ops** | $I(ids, c)$ and $D(id)$; $c$ is a character | | | | | |
| **Object** | <idc,idp,idn, v,c> **26B** | <idc, deg, v,c> **14B** | <idc,deg,nlink,hlink, v,c> **22B** | <idc,idp,nlink, hlink, v,c> **42B** | <id> an *object* has a variable size bounded by $C$ | |
| **Identifier (id)** | <sid, seq> **8B** | | | <sn,sum,sid,seq> **16B** | <i₁,sid>,... <iₙ,sid>, seq each *id* has a variable length bounded by $C$ | <i₁,sid,seq>... <iₙ,sid,seq> |
| **Identifiers (ids)** | <idp,idc,idn> **24B** | <idp, idc, idn, deg> **28B** | | <idc, idp> **32B** | NA | |
| **Identifier ordering** | *All totally ordered* | | | | | |
| | without additional constraint | | | + causal order | + positional order | |
| **Timestamp** | Scalar-based | | | Vector-based | NA (but require causally ordered broadcast) | |
| **Time complexity** | $O(C_t^3)$ | $O(C_t^2)$ | $O(C_t^2)$ | $O(C)$ with G.C. | $O(C)$ for local operations; $O(C \circ log(C))$ for remote ops | |
| **Space complexity** | $O(C_t)$ | | | $O(C)$ with G.C. | between $O(C)$ and $O(C^2)$ | |

With this extra tree, together with its additional complication and costs, the time complexity of this search process could be reduced to $O(log(C_t))$ or $O(log(C))$, which improves the upstream execution time by *"several orders of magnitude"* as claimed in [9]. For comparison, OT solutions commonly have the complexity $O(1)$ for the whole process of handling a local operation.

*4.1.5 Space Complexity $O(C_t)$ or between $O(C)$ and $O(C^2)$*. Apart from time complexities, space complexities of CRDT solutions are also characterized by $C_t$ or $C$. The internal object sequence has the space complexity $O(C_t)$ for CRDT solutions with tombstones (e.g. WOOT variations), or between $O(C)$ and $O(C^2)$ for those without tombstones (e.g. Logoot [84]).

Last but not least, the actual size of each object in the sequence deserves attention too: for each character of one byte in a text document (most CRDT solutions are confined in plain text co-editing), the corresponding internal object may have a size of 14, 22, or 26 bytes for WOOT variations, 42 bytes for RGA, or a variable size (bounded by $C$) for Logoot variations (see Table 3). This means that the space overhead of a CRDT object sequence is at least 14 to 42 times larger than the original size of the external text document, without counting tombstones.

## 4.2 Tombstone Overhead Issues in WOOT Variations and RGA

WOOT variations and RGA are different from each other in concrete representations of internal object sequences and operations, but they have one thing in common: the use of tombstones to represent deleted objects in the object sequence. The very need for tombstones was to support identifier-based search of the object sequence in the face of concurrent operations. The major problem with tombstones is the time and space overhead. The following is a telling story of the unacceptable cost caused by tombstones, reported directly by WOOT researchers [84]:

> "*For the most edited pages of Wikipedia, the tombstone storage overhead can represent several hundred times the document size. Tombstones are also responsible of performance degradation. Indeed, in all published approaches, the execution time of modification integration depends on the whole document size – including tombstones*" (page 1 of [84]). ... "*While the 'George W. Bush' page contains only about 553 lines, the number* *of deletions is about 1.6 million. As a consequence, tombstones-based systems are not well-suited for such documents since we obtain 1.6 million tombstones for only 553 lines*" (page 5 of [84]).

While the prohibitive tombstone cost ($C_t$ = $C$ × *2894* in this example) quoted above is only one case study and not necessarily generalizable, the tombstone overhead is indeed a major and inherent issue with WOOT variations and RGA.

Various attempts were made trying to remove tombstone garbage. In RGA [48], a garbage collection scheme was proposed, which was based on the following conditions[8]: a tombstone object can be removed *only if*: (1) no future operation will be concurrent with the *delete* operation that converted the object into a tombstone (similar to the operation garbage collection condition based on vector-clocks in GOT [60]); (2) no future operation will have a total order that is earlier than the total order of the neighboring object immediately after the tombstone in the sequence, which is an additional condition due to RGA's special way of maintaining and using tombstones.

However, WOOT, WOOTO and WOOTH have no garbage collection schemes of their own, and also cannot adopt the RGA garbage collection scheme due to the absence of vector-clocks and different ways of maintaining and using tombstones and neighboring objects under complex concurrent co-editing scenarios [4,44,85].

A more recent tombstone-based CRDT solution, named Yjs [40], is noteworthy. Yjs is special in its extensions for supporting rich-text co-editing; but its core is based on WOOT with variations to reduce time and space time complexity. Yjs claims to have two major improvements over WOOT: (1) the time complexity of $O(C_t)$ (in contrast to $O(C_t^3)$ of WOOT); and (2) a tombstone garbage collection scheme (WOOT does not have one) without using vector-clocks (in contrast to RGA). Unfortunately, those Yjs variations incurred correctness issues (e.g. a tombstone is collected as garbage "*after a fixed time period*" [40], which is clearly incorrect) due to the intricacies in maintaining tombstones and neighboring objects under complex concurrent scenarios (also see Section 4.1.1). Yjs rich-text extensions are examined and discussed in [78].

### 4.3  Identifier Length and Space Explosion Issues with Logoot

As the tombstone issue is clearly a road blocker for WOOT variations to be applicable to real world co-editors, an alternative CRDT solution, named Logoot [84], was proposed to avoid using tombstones in the object sequence. However, this CRDT solution imposed a special *positional ordering* constraint on identifiers: object identifiers must have a total ordering that is consistent with the positional ordering of the corresponding objects in the object sequence, which mirrors the sequence of characters in the external document state.

As shown in Table 3, all CRDT solutions, with or without tombstones in object sequences, have imposed a total ordering on identifiers for various purposes, e.g. to determine the location of an *insert* operation when multiple concurrent operations are inserting at the same location in the object sequence (for WOOT variations and RGA). The positional ordering constraint for the Logoot (and its variations) identifier is quite special, which is to determine the location of an operation (being *insert* or *delete*, concurrent or sequential) in the object sequence (at remote sites) ─ the key to Logoot's correctness and complexity, and hence worth further elaboration.

For any two objects A and B in the object sequence, with identifiers $id_a$ and $id_b$, respectively, if A is positioned before B in the object sequence, $id_a$ must be ordered before $id_b$ in the identifier ordering as well. This property is essential for searching target objects or locations in the object sequence, without the help of tombstones, in the face of concurrent editing. For Logoot to work, the key is to devise identifiers that possess *uniqueness*, *immutability* and *positional ordering* properties, which turned out to be (and remain) a major challenge.

The basic idea in the Logoot identifier scheme is to assign each object with an integer in a number system of a chosen base (e.g. $2^{64}$) according to the object's position in the object sequence.

---

[8] It is noteworthy that the key for RGA to improve WOOT and its variations (in time complexity and garbage collection) was due to the use of vector clocks and causal-ordering, which were also adopted in many OT solutions, but commonly dismissed by CRDTs (e.g. WOOT) on the ground of being non-scalable and unsuitable for peer-to-peer co-editing [78].

For any two objects A and B, if A precedes B in the object sequence, the integer assigned to A must be smaller than that to B. To insert a new object X between two existing objects A and B with integers $p$ and $q$ in their identifiers, respectively, the identifier scheme will assign object X with an integer $i$, which is randomly taken between $p$ and $q$, such that $p < i < q$.

One issue with this basic scheme stems from the nature of concurrent editing: what if two users are inserting concurrently between the same pair of objects and the identifier schemes at the two co-editing sites generate the same random integer $i$? To break the *concurrent-insert-tie*, Logoot coupled the random integer with a site identifier (*sid*) to make a tuple *<i, sid>*, and use *sid* to break a tie. This extension, however, was still not enough.

Another issue with the identifier scheme is related to the limited space between any two integers: what if a user inserts a large number of objects between two existing objects, and the required integers for identifying those objects exceed the available integers between the identifiers of neighboring objects? This led to further extensions of the Logoot identifier from a single tuple to a variable number of tuples, bounded by the big *C*, and an additional local operation sequence number (*seq*) was added to maintain *uniqueness* (see Table 3).

In the end, Logoot got a variable and potentially lengthy identifier scheme, which could lead to high space and time costs in object sequence representation and searching [84]. Average lengths of identifiers depend on editing patterns, and the worst case complexity of Logoot identifiers is $O(C)$ [84,86] – to put this in plain terms: the length of a Logoot identifier to represent a single character is proportional to the length of the document, which led to the space complexity $O(C^2)$ for the object sequence. The time complexity for handling a local insert is $O(C)$, but $O(C \circ log(C))$ for a remote insert [4,84,86]. The space and time costs in Logoot identifiers turned out to undermine what Logoot was originally proposed to achieve, i.e. to avoid time and space costs caused by tombstones in WOOT variations.

To get the identifier length and object sequence space explosion issues under control, various patches were added to the Logoot identifier scheme. Those patches, unfortunately, led to increasingly complicated schemes, which can generate identifiers violating required properties, and manifest as inconsistencies in document states, as revealed in the following sections.

### 4.4 Hidden Correctness Issues in Logoot

In this section, we present multiple correctness issues in Logoot that we discovered as part of this investigation. In Fig. 2, we use a simple scenario with three operations to illustrate these issues. Like in Fig.1, we use ES for External State (visible to users), EO for External Operation (generated from ES by users), and IS for Internal State (the object sequence manipulated by Logoot).

In the following examples, we use the identifier generation scheme described in [84,86] (two versions are basically the same). For simplicity, we use 10 as the integer base for identifiers, MAX as a symbolic value larger than the biggest integer 9 in the base, and <0, NA,NA> and <MAX,NA,NA>, which were also used in [86], as two special identifiers marking the start and end of the object sequence. Initially (the portion above the first horizontal dashed line in Fig. 2), the external state contains two characters "XY", which correspond to the two internal objects with identifiers <1,1,1> and <3,1,2>, respectively. In each tuple of an identifier, the first number is an integer, the second number is the site identifier, and the third number is the sequence number.

*4.4.1 Concurrent-Insert-Interleaving Puzzle.* Among all issues with the Logoot identifier scheme, the first critical one is the *random interleaving* anomaly, which could occur whenever two users concurrently insert continuous characters between the same pair of existing characters.

As shown in Fig. 2, User A generates $EO_1 = I(1, \text{"ab"})$ to insert two characters "ab" between "X" and "Y"; and concurrently User B generates $EO_2 = I(1, \text{"AB"})$ to insert two characters "AB" between "X" and "Y". Internally, Logoot at the User A site first uses the position 1 in $EO_1$ to find the insert location and obtain two neighboring identifiers <1,1,1> and <3,1,2> in $IS_{10}$. Since there exists only one available integer between 1 and 3 (in the first tuples of the two neighboring identifiers), a new tuple has to be added to generate two new identifiers for the two inserted characters. According to the identifier generation scheme in [84,86], a range of legitimate identifiers (shown in **Box A** in Fig. 2) are available for random choices. Without losing generality, we choose two identifiers <1,1,1><9,1,3> and <2,1,4><8,1,4> from this range to identify the two inserted characters. Then, Logoot inserts these two identifiers at the location 1 in the object sequence. Afterwards, two identifier-based operations are propagated to User B. At the User B site, a similar process occurs, two identifiers <1,1,1><9,2,1> and <2,2,2><0,2,2> are generated (from **Box B**) to identify the two characters, and two identifier-based operations are propagated to User A.

After receiving identifier-based operations from a remote site, Logoot uses the identifiers in remote operations to determine the insert locations in the object sequence. After applying the identifier-based concurrent operations at each other sites, the combined final (both internal and external) states are consistent, but the four characters and their corresponding internal identifiers are *interleaved*: the external document state becomes "XaAbBY", rather than "XabABY" (or "XABabY"), which would normally be expected by users.

It should be highlighted that the four identifiers are legitimate identifiers (according to [84,86]), and meet the *uniqueness*, *immutability* and *position-ordering* properties required by Logoot, but their orders are randomly interleaved. This anomaly is rooted in Logoot's fundamental random positional identifier generation scheme.

*4.4.2 Inconsistent-Position-Integer-Ordering Puzzle.* Before the second horizontal dashed line in Fig. 2, the concurrent-insert-interleaving abnormally has manifested itself, but there is another hidden problem inside the internal states (both $IS_{12}$ and $IS_{22}$): the two adjacent identifiers $id_1$ = <2,1,4><8,1,4> and $id_2$ = <2,2,2><0,2,2> have an *inconsistent-position-integer-ordering* problem, meaning their positional order ($id_1 < id_2$) is inconsistent with their integer order (28 > 20). This hidden problem would manifest as an *infinite loop* in the Logoot identifier generation scheme when any user inserts a character between these two adjacent objects. For example, when User A generates $EO_3$ = $I$(4, "cd") to insert "cd" at position 4, i.e. between "b" and "B" in $ES_{12}$, as shown in Fig. 2, Logoot would run into an infinite loop and fail to generate identifiers.

In general, the Logoot identifier generation scheme will run into an infinite loop and fail to generate any new identifier between two neighboring identifiers with position integers $p$ (for *left* identifier) and $q$ (for *right* identifier) if $p \geq q$ (in other words, their position and integer orders are inconsistent), which could occur when two users are inserting at the same location concurrently.

The infinite loop flaw was found in the first Logoot paper [84], remained the same in a late version [86], and never corrected in late CRDT publications. However, we found several patches introduced to avoid the infinite loop in open source codes[9,10], which implemented the Logoot identifier scheme. Nevertheless, none of those patches could really solve the problem without running into new problems, which are illustrated in the next subsection.

*4.4.3 Position-Order-Violation Puzzle.* We use the patch in the Logoot library (referred to in footnote 9) implemented by Logoot authors [9], to illustrate the position-order-violation puzzle, which is caused by the patches introduced to resolve the infinite loop flaw, as shown under the second horizontal dashed line in Fig. 2.

When User A generates $EO_3$ = I(4, "cd") to insert two characters "cd" between "b" and "B", the Logoot library would avoid the infinite loop by changing the *right* neighbor identifier from <**2**,2,2><0,2,2> into <**3**,2,2><0,2,2> (only inside the identifier generation algorithm, not in the internal state), to allow the use of a range of *illegitimate* identifiers from <2,1,4><8,1,4><1,1,*> to <2,1,4><0,2,2><9,1,*>, shown in **Box C** in Fig. 2. However, this patch could cause numerous abnormal cases that cannot be handled by the original identifier generation scheme (e.g. the *ConstructId* function in [86]), which in turn requires additional patches to deal with. For example, the tuple <0,2,2> in identifiers <2,1,4><0,2,2><*,1,*> is inherited from the corresponding tuple in the right neighboring identifier <**3**,2,2><0,2,2>, and this inheritance is forced by one patch in the Logoot library[11]. Unfortunately, the position-order-violation problem manifests itself among these illegitimate identifiers, e.g. between the two identifiers <2,1,4><8,1,4><9,1,5> and <2,1,4><0,2,2><4,1,6> (and among other pairs as well) in this range. Trouble would occur when these two identifiers are assigned to the two new objects for the two characters "cd".

---

[9] https://github.com/coast-team/replication-benchmarker.
[10] https://github.com/rudi-c/alchemy-book. It is worth noting that the author of this work also detected various issues in Logoot and pointed out Logoot "*missing what I think are key details on how to handle certain edge cases*", and devised his own patches to deal with those key missing details. Those patches also came with problems which may result in state inconsistencies or system crashes. Detailed analysis of those issues is beyond the scope of this article.

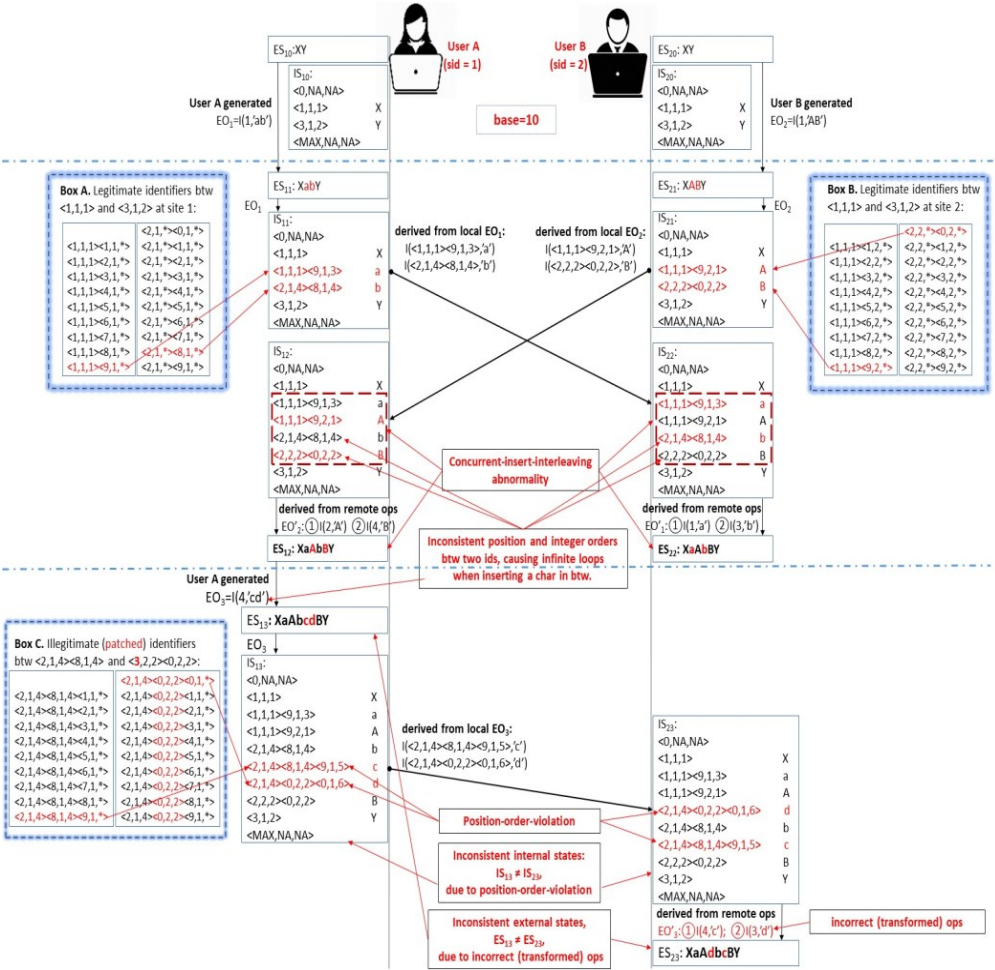

Fig. 2. Illustration of multiple correctness issues in Logoot, including *concurrent-insert-interleaving*, *inconsistent-position-integer-ordering* (and *infinite loop* flaw), and *position-order-violation*.

This position-order-violation does not immediately cause a trouble at the local site as the local insertion position is determined by the position number 4, rather than by these identifiers; the trouble occurs when Logoot at the remote User B site uses these identifiers to determine their positions in $IS_{22}$, and inserts "c" and "d" at corresponding positions, which results in inconsistent internal states, i.e., $IS_{13} \neq IS_{23}$, which in turn leads to incorrect external (transformed) operation $EO_3'$, which finally results in inconsistent external states, i.e. $ES_{13} \neq ES_{23}$, as shown in Fig. 2.

It should be pointed out that the inconsistent-position-integer-ordering problem (and the associated infinite loop flaw), and the position-order-violation puzzle could occur under numerous circumstances, e.g. the reader can use different identifier combinations in **Boxes A, B** and **C** in Fig. 2 to create varieties of similar puzzles.

---

[11] This patch is implemented in functions *generateLineIdentifiers* and *constructIdentifier* in https://github.com/coast-team/replication-benchmarker/blob/master/src/main/java/jbenchmarker/logoot/.

It remains a critical issue for Logoot to find a correct identifier scheme. Logoot variations, such as LogootSplit [5] and LogootUndo [86], tried to extend Logoot from supporting character-wise to string-wise operations and from supporting *do* to *undo*. Unfortunately, new identifier schemes for string operations and *undo* were even more complicated than that for character-wise operations and *do*-only Logoot solutions, and their correctness was not verified either.

### 4.5 General Correctness Issues with CRDT for Co-Editors

In general, the correctness of key components in CRDT solutions, e.g. object identifiers, object sequences, object sequence searching and manipulation schemes, remains to be verified, using well-defined criteria, which are yet to be established as well. In contrast, representative OT solutions have their key components, i.e. control algorithms and transformation functions, theoretically verified under well-established algorithmic correctness criteria (see "*the multi-facets of OT correctness*" in [66]) − context-based conditions and transformation properties (discussed in Section 3), and experimentally validated in working co-editors based on these solutions (see [78]).

Consistency (e.g. convergence) claims of a co-editing technical solution cannot be derived from any theory that assumes components of this solution will work according to theoretic requirements. For example, Logoot identifiers are required to possess the positional ordering property − one necessary condition for Logoot to achieve convergence, but the designed Logoot identifier schemes actually violated this required property and led to inconsistency, and even run into infinite loops and failed to generate any identifier, as illustrated in Fig. 2. Claiming "*we know exactly why CRDTs converge*" (see footnote 4) can neither eliminate those CRDT algorithmic flaws, nor help guarantee the correctness of any CRDT solutions for co-editors.

### 4.6 Summary of Correctness and Complexity of CRDT Solutions

One main motivation of the first CRDT solution (WOOT) was to address the FT puzzle and the CP2-violation issue in OT for plain-text co-editing, which have been solved under the OT approach in numerous ways (see Section 3.3). As an OT-alternative in co-editing and following the same general transformation (GT) approach (see Section 2), CRDT solutions did the job without using OT algorithms, but with CRDT-special object sequences, identifier-based operations, and schemes for manipulating such sequences and operations, which came with CRDT-special correctness and complexity issues (see time and space complexities of representative CRDT solutions in Table 3).

Some CRDT issues (e.g. tombstone overhead for WOOT variations, inconsistent-position-integer-ordering and position-order-violation for Logoot variations, etc.) are still open for resolution, but some others (e.g. big $C/C_t$ complexities in time and space for all CRDT solutions, and concurrent-insert-interleaving for Logoot variations, etc.) are inherent to the basic approaches taken by those CRDT solutions, and it is unclear whether they can be solved without fundamental changes to the basic approaches.

## 5 DISCUSSION ON CORRECTNESS AND COMPLEXITY OF OT AND CRDT

Based on the results reported in previous sections, we further examine major myths and misconceptions surrounding correctness and complexity of OT and CRDT in this section.

Historically, OT predated CRDT for over one decade. CRDT was proposed to address the same consistency maintenance issues (mostly confined in plain-text co-editing) and to meet the same consistency requirements as OT does, as discussed in Section 2. Later proposals ought to compare with prior art for justification. To justify CRDT in the presence of the OT prior art, CRDT articles often handwave messages like: "*most OT algorithms have been proved incorrect*" (e.g. see footnote 4 and examination in Section 5.1), "*the only OT algorithm proved correct is TTF*" (e.g. see footnote 4 and examination in Section 5.2), and "*CRDT algorithms is more efficient by a factor up to 1000 compared to representative OT algorithms*" (e.g. see footnote 4 and examination in Section 5.3), by quoting a few earlier publications (e.g. [4,24,45], examined in Section 5.1) without independent validation. Later CRDTs were often justified by comparing with prior CRDTs only, as if OT had been ruled out in the "*post*-OT" era, resulting in numerous self-justified CRDT variations (see examples in Section 4).

These fallacious messages had the effects of distorting the reality, causing major confusions in co-editing communities, particularly to new researchers and practitioners, and hindering the progress of co-editing research and application. Refuting fallacies with facts about correctness and complexity of OT and CRDT is necessary and important to advancing co-editing research and real world application.

### 5.1 Misconceptions and Facts about OT Correctness

A large body of CRDT literature in co-editing has been built on the notion that *OT is incorrect*. This notion has served as a primary justification for CRDT (e.g. see footnote 4): if OT were indeed incorrect, then any *un*-OT solution, such as CRDT, could potentially be better off and even justifiable to bear high costs (e.g. the big $C/C_t$–determined complexities and tombstone overhead, revealed in Section 4), or to have hidden algorithmic flaws (e.g. Logoot flaws illustrated in Fig. 2). Nevertheless, this notion is fundamentally wrong as it is based on misconceptions and contradicted by basic facts about OT correctness, as elaborated below.

*5.1.1 The CP2 Syndrome.* Achieving convergence is one key requirement for OT and any consistency maintenance solution to co-editing. Two transformation properties CP1 and CP2 (see details in Section 3.2.1) are directly related to achieving convergence in OT [15,59,60].

In [46], one theorem established that CP1 and CP2 [12] are two necessary and sufficient conditions to achieve convergence under the adOPTed control algorithm[13], which was proposed as a solution to the dOPT puzzle [59]. This theorem is important and applicable to transformation functions in OT solutions, such as adOPTed, that allow concurrent operations to be transformed in arbitrary orders, more precisely, under different contexts [72]. Unfortunately, this theorem was often misinterpreted – a source of common misconceptions surrounding OT correctness, particularly about CP2 correctness, which are collectively referred to as the *CP2 syndrome* below.

*5.1.2 Are CP1 and CP2 Necessary and Sufficient Conditions for OT Correctness?* The top symptom of the CP2 syndrome is to misinterpret CP1 and CP2 as necessary and sufficient conditions for the correctness of all transformation functions or even an OT solution as a whole. This interpretation had distorted the real meaning and importance of CP1 and CP2, and misled many to take CP1 and CP2 as two golden rules in evaluating OT correctness, resulting in numerous flawed arguments and claims in co-editing literature and forum discussions among practitioners.

In fact, CP1 and CP2 are neither necessary nor sufficient for the correctness of transformation functions, let alone for a whole OT solution. CP1 and CP2 are *unnecessary* for transformation functions because they can be avoided when an OT solution has used a suitable OT control algorithm (e.g. CP1-avoidance algorithms like GOT [60] and TIBOT [29,89]; numerous CP2-avoidance algorithms discussed in Section 3.3). CP1 and CP2 are *insufficient* because they govern convergence only, but not other consistency requirements for co-editors, such as causality-preservation and, most importantly, intention preservation [60,75,76] (also see Section 2).

Without the intention preservation requirement, transformation functions can preserve CP1 and CP2 trivially. For example, consider the function: *trivial-TF*($O$, $O_x$) = $O'$, which transforms $O$ against $O_x$ to produce $O'$, where $O'$ is an operation that always replaces existing contents of the document with a number $X$. It can be shown that this *trivial-TF* preserves CP1 and CP2 and always ensure convergence. By assigning $X$ to an arbitrary number, one can get an infinite number of transformation functions capable of preserving CP1 and CP2, but none of them is meaningful, let alone be correct for co-editing.

*5.1.3 Is CP2-Violation the Root of All Inconsistency Problems?* Another symptom of the CP2 syndrome is to exaggerate the role of CP2 in OT correctness by attributing every divergence puzzle to CP2-violation. CP2-violation will certainly result in divergence, but not all divergences are caused by CP2-violation. Divergence is only a *symptom* of a puzzle and could be *caused* by other factors unrelated to CP2 (or CP1) [66]. Accurate attribution of an OT puzzle to the true root transformation condition is crucial not only to resolving the puzzle, but also to evaluating the correctness of OT in general.

---

[12] In [46], CP1 and CP2 are named as TP1 (Transformation Property 1) and TP2 (Transformation Property 1), respectively.
[13] The adOPTed algorithm can resolve the dOPT puzzle due to its capability of ensuring the context-equivalence condition [59], rather than requiring functions to preserve CP1 and CP2. However, the context-equivalence condition was not explicitly stated in [46], but implied in the description of adOPTed, whereas CP1 and CP2 were explicitly stated in a theorem, which was often misinterpreted as the reason for adOPTed to be able to resolve the dOPT puzzle.

For example, the famous dOPT puzzle had resulted in divergent states, and it was due to violation of the context-equivalence condition (responsible by control algorithms), rather than violation of CP2 (responsible by transformation functions). Discovering the context-equivalence condition was crucial in resolving the dOPT puzzle [59]. Unfortunately, after the dOPT puzzle had been resolved for over one decade, we still see co-editing publications (e.g. [45,48]) trying to attribute this puzzle to CP2-violation and to relate proposed TTF (*Tombstone Transformation Functions*) [45] to the control algorithm puzzle. This example revealed a glaring misunderstanding of basic OT correctness conditions and relationships between control algorithms and transformation functions, and also illustrated a phenomenon in co-editing literature, in which the importance of CP2-related work was grossly inflated.

*5.1.4 Are Most OT Solutions Incorrect with Respect to CP2?* Yet another and maybe the most common symptom is to dismiss most (or even all) OT solutions for not preserving CP2 (e.g. [4,5,9, 24,43,45,48,86]). Arguments along this line could be traced back to the early history of exploring alternative solutions to the FT puzzle (a case of CP2-violation).

After the discovery of the FT puzzle [58,60], numerous attempts were made to resolve this puzzle, resulting in a large number of proposals [24,25,30,31,32,33,43,44,45,57]. Though different from each other, those works share some common characteristics. One was to magnify the importance of CP2 and CP2-related work in relation to OT correctness (see Sections 5.1.2 and 5.1.3). Another phenomena, often "justified" by the inflated CP2 importance, was to make radical changes to various core OT components for the sake of addressing CP2, e.g. to change underlying data and operation models (e.g. tombstone-based TTF [45]), or the basic strategy of separating generic control algorithms and application-specific transformation functions (see footnote 5), which are fundamental to OT but have little to do with the FT puzzle. Consequently, those changes often brought in new and unexpected issues, in both correctness and complexity, which were far more complicated than the original FT problem they were proposed to address.

There were some published works claiming to have disproved all prior (by then) OT solutions in terms of CP2-correctness (e.g. using theorem provers [24,25]), and proposed new solutions that were proven to be CP2-correct (using the same theorem provers); but those proposals or proofs were repeatedly found to be flawed later (e.g. see counter-examples reported in [24,30,32,33,57]). Erroneous results and misconceptions generated from those attempts had the effect of creating the myth that OT was full of puzzles spiraling out of control, which had caused major confusions among practitioners and later researchers entering the field.

*5.1.5 Myths in CP1 and CP2 Verification.* Another related myth is that verifying CP1 and CP2 is an exponential explosion problem, which is too complex and "even impossible" for manual checking [8,24,25]; so tools like theorem-provers and model checkers had to be used for this purpose. As reported in [8], hundreds of thousands of states had to be checked in order to verify a co-editing scenario with a few operations (e.g. 331,776 states for merely 4 operations), which was beyond the capacity of the adopted model checker tool. Nevertheless, the state explosion problem is not inherent to OT verification itself, but caused by adopted verification methods (e.g. [8]), in which concurrency relationships (taken care of by control algorithms) and positional relationships (responsible by transformation functions) among operations were mixed, rather than separated from each other ─ a common pitfall in those OT verification papers.

By separating control algorithm verification (based on concurrency and context conditions) from transformation function verification (using criteria CP1 and CP2, and combined-effects, etc) [34], one verification method in [75] exhaustively covers all distinctive transformation cases (a total of 9 cases for verifying CP1 and combined-effects, and 58 cases for CP2 verification), which can be checked even manually, under the same character-wise operation and data models as [8]. In [76], the verification method in [75] has been extended to cover all transformation cases under a more general string-operation model (which needs 14 cases for verifying CP1 and combined-effects, and 422 cases for CP2 verification). A software tool, named OTX (OTXplorer) [69], was developed to automate the process of generating and checking all possible transformation cases. OTX has also been used to support design, experimentation, and verification of different string-wise transformation functions. One set of concrete transformation functions, capable of preserving CP1 (plus commonly adopted combine-effects) and CP2 and verified by using OTX, can be found in Table 3 of [76].

*5.1.6 Facts about the FT Puzzle and OT Correctness with Respect to CP2.* To help clear up these myths and misconceptions, we highlight the following basic facts.

1. Despite a variety of CP2-violation puzzles or counter-examples reported in literature (see Section 5.1.4), all reported puzzles were either variations of the same basic FT puzzle reported in [58,60], or they were simply derivatives of erroneous solutions proposed to solve the original FT puzzle. In other words, those works did not make any new puzzle discovery in prior existing OT solutions [66,75], but only introduced new errors in later proposed FT solutions.
2. All those attempts (covered in Section 5.1.4) had been confined to a primitive model with character-wise *insert* and *delete* operations in text editing. Based on exhaustive examination of all possible transformation cases under this primitive model (see Section 5.1.5), it has been proven that the FT puzzle is the only possible CP2-violation case in OT solutions that support commonly adopted combined-effects for pair-wise concurrent operations in text editing [75,76]. This result explains why those attempts (see the prior point 1) could not make any new puzzle discovery under this basic model.
3. Last and the most important fact: all possible CP2-violation cases under a more general string-wise operation model[14] for text editing have been detected and solved by numerous solutions based on both CP2-preservation (applicable to text editing [76]) and CP2-avoidance (applicable beyond text editing) strategies (see Section 3), which form the bases for building today's working co-editors and industry products [78].

In a nutshell, the notion that "*most OT algorithms have been proved incorrect*" (with respect to CP2 or as a whole) is plainly groundless and false. Inevitably, those works that were built on such a false notion are groundless as well.

## 5.2  Twin Solutions to CP2-Violation: WOOT and TTF

Among numerous alternative proposals (other than the ones summarized in Section 3.3) to address the CP2-violation issue, a pair of solutions are particularly noteworthy: one is WOOT, and the other is TTF (*Tombstone Transformation Functions*), both of which were based on the same idea of internal object sequences with tombstones, and proposed at nearly the same time by the same authors [43,44,45].

*5.2.1 What is Special about WOOT and TTF?* Basically, WOOT was proposed as an OT alternative capable of avoiding CP2-violation by introducing internal object sequences with tombstones and identifier-based operations, without using OT (and originally without the label of CRDT as well); the TTF solution was proposed as a solution capable of preserving CP2 by introducing internal object sequences with tombstones (similar to WOOT), and designing transformation functions based on the new internal object sequence. What made WOOT and TTF special was not so much their alternative ways of dealing with the CP2 issue, but their unjustified claims and roles in late development: WOOT was late named as the *first* CRDT and mystified to possess a capability of making concurrent operations natively commutative, among other superiorities over OT [4,5,9,28,35,42,43,44,48,50,51,84,85,86]; TTF was claimed to be the *first* and often cited as the *sole* correct OT capable of achieving CP2 correctness [4,5,9,24,43,45, 48,86] (also see footnote 4).

---

[14] In [76], we detected and resolved an additional *False-Border* (FB) puzzle under a pair of string-wise insert and delete operations, and had shown that the FT and FB puzzles were the only two possible CP2-violation cases in OT solutions supporting string-wise operations (with commonly adopted combined effects of concurrent operations) in text co-editors. The work in [76] also re-confirmed FT is the only puzzle under the character-wise operation model.

*5.2.2 Using TTF as the Sole Correct OT for Comparison with CRDT.* Being claimed as the sole correct OT solution, TTF was often used by CRDT advocates as the OT representative in comparison with later CRDT solutions. Quite some claims about CRDT superiority over OT were based on the comparison between CRDT solutions (e.g. Logoot, RGA, etc.) and TTF (typically integrated with the SOCT2 control algorithm [56]). For example, the TTF solution was reported to be outperformed by Logoot and RGA for a factor up to 1000 in [4]. This *1000-times-gain* claim was highly remarkable and widely cited as an experimental evidence for CRDT's performance superiority over OT [4,5,9,48,86] (also see footnote 4).

It would be interesting to validate whether and how Logoot and RGA had truly achieved *1000-time-gain* over TTF (+SOCT2)[15], but that is outside the scope of this paper. What we want to point out here is that those CRDT claims over OT are groundless and false, because: (1) they are contradicted by the facts that numerous OT solutions have been proven to be correct with respect to well-established conditions and properties (including CP1 and CP2) *before* and *after* WOOT and TTF appeared (see Section 3); and (2) they are also mistaken about what TTF really is: TTF is not a typical OT solution but a hybrid of CRDT and OT, or "*a CRDT in disguise*" (see footnote 4).

*5.2.3 What is TTF Really?* In the TTF solution, an internal tombstone-based object sequence is maintained, which is a characteristic CRDT component (like WOOT). Another component is a set of special CP2-preserving transformation functions (i.e. TTF [45], defined for character-wise *insert* and *delete* operations on a sequence of objects with tombstones), which are integrated with an OT control algorithm (e.g. SOCT2 [56]) to transform operations. The latter component is similar to OT, with one subtle but crucial difference: operations being transformed by TTF are not editing operations defined on external document as in typical OT solutions (discussed in Section 3), but internal operations which are defined on and only applicable to the internal tombstone-based object sequence. Consequently, additional conversions between internal and external operations are required, which is typical to CRDTs (like WOOT as illustrated in Fig.1).

Due to its hybrid nature, the TTF solution bears the costs of both CRDT and OT, with the main costs dominated by its CRDT components. The OT control algorithm SOCT2, often used as the TTF companion, has the time complexity of $O(c^2)$, which is significantly lower than $O(C_t^3)$ of WOOT; the costs of TTF CRDT components include the maintenance of the tombstone-based object sequence and associated search schemes (each with the time complexity $O(C_t)$, where $C_t$ is multiple orders of magnitude larger than $c$ and even $c^2$) for conversions between internal and external operations. We refrain from detailed comparison of TTF with typical OT or CRDT in this paper, but will do comprehensive comparisons of OT, CRDT, TTF, and other alternatives (e.g. [21,30,31,32,33]), that are based on the same general transformation approach in a future paper[16].

## 5.3 Comparison of OT and CRDT in Time and Space Complexity

In addition to the *1000-times-gain* claim over OT (based on reported comparison with TTF), CRDT has claimed general superiority over OT in theoretic time and space complexity. CRDT complexity claims are secondary to correctness claims and were often cited when one was not contented with or uncertain about dismissing OT on the ground of correctness. After all, if OT were indeed incorrect, what is the point to argue about complexity differences between a faulty OT and a correct CRDT (see footnote 4)?

---

[15] It is worth noting that the performance results reported in [4] did not cover WOOT for the reason "*it is obviously outperformed by its optimized versions WOOTO and WOOTTH algorithms*", and did not have a comparison of WOOT with TTF+SOCT2 (used as the OT representative) either. Given the significant differences between $O(C_t^3)$ (for WOOT) and $O(c^2)$ (for SOCT2) and between $C_t$ and $c$, it is reasonable to derive that WOOT could be outperformed by TTF+SOCT2, and, by deduction, outperformed by Logoot and RAG for a factor of more than 1000 as well, according to [4].
[16] We have used the GT framework to describe TTF and show its CRDT and OT hybrid nature in Section 4.3 of [78].

*5.3.1 Complexity in Theory.* Unfortunately, not only CRDT correctness claims are untrue, but CRDT complexity claims are false as well, which have been elaborated in detail in Sections 3 and Section 4. For easy comparison, we summarize complexities of representative OT and CRDT solutions in Table 4, which clearly disprove CRDT claims in complexity, as highlighted below:

1. For local operation processing:
   a. all OT solutions have a constant time complexity $O(1)$;
   b. the best time complexity for CRDT is $O(C)$ for RGA. In [9], one extra binary tree was proposed to speed up searching the internal object sequence, which could reduce the theoretic time complexity to $O(log(C_t))$ or $O(log(C))$, at additional costs for managing the extra tree (see discussions in Section 4.1.4).
2. For remote operation processing, the best time complexities for both OT and CRDT are linear with respect to $c$ or $C/C_t$:
   a. $O(c)$ for OT (e.g. POT, COT, Google OT, Jupiter, NICE, TIBOT, and SOCT4).
   b. $O(C)$ for RGA.
3. The best space complexities for both OT and CRDT are linear with respect to $c$ or $C/C_t$:
   a. $O(c)$ for OT (e.g. Google OT, Jupiter, NICE, TIBOT, and SOCT4);
   b. $O(C)$ for RGA, or $O(C_t)$ for WOOT variations.

CRDTs clearly have higher time complexities than OT in local operation processing. In terms of time complexity for remote processing and space complexity, both CRDT and OT have linear complexities under respective variables $c$ (for OT) and $C/C_t$ (for CRDTs), which are seemingly similar in theory, but significantly different in practice (see Section 5.3.2 below).

Table 4 Space and time complexities of representative OT and CRDT solutions.
In real-time co-editing sessions, m is the number of users, usually m ≤ 5.

|  | OT | CRDT | |
|---|---|---|---|
|  | **TIBOT** [29], **Jupiter** [39], **adOPTed** [46], **NICE**[52], **SOCT2**[56], **GOTO**[59], **GOT**[60], **COT**[72], **SOCT4**[82], **Google OT** [83], **POT**[89] | **Tombstone-based WOOT variations** [4,44,85] + **RGA** [48] | **Non-tombstone-based Logoot variations** [5,84,86] |
| **Space** | $O(c)$ for Google OT, Jupiter, NICE, TIBOT, SOCT4, $O(c \cdot m)$ for POT, GOTO, GOT, SOCT2 $O(c \cdot m^2)$ for COT, adOPTed, | $O(C_t)$ for WOOT variations $O(C)$ for RGA | $O(C)$ to $O(C^2)$ for all |
| **Time** | **Local**: $O(1)$ for all | **Local & remote:** $O(C)$ for RGA $O(C_t^2)$ for WOOTO&WOOTH $O(C_t^3)$ for WOOT **Local:** $O(log(C_t))/O(log(C))$ [9] | **Local:** $O(C)$ for all **Remote:** $O(C \circ log(C))$ for all |
|  | **Remote:** $O(c)$ for Google OT, POT, COT, TIBOT, Jupiter, NICE, SOCT4 $O(c^2)$ for adOPTed, GOTO, GOT, SOCT2 | | |

*5.3.2 Complexity in Practice.* The real complexity differences between OT and CRDT solutions should be examined not only by theoretic expressions using the big-$O$ notation, but also by practical evaluation of the input variables in those theoretic expressions. OT complexity variable $c$ is often bounded by a small value, e.g. $0 \leq c \leq 10$, for real-time sessions with a few (≤ 5) users. CRDT complexity variable $C$ is typically orders of magnitude larger than $c$, e.g. $10^3 \leq C \leq 10^6$, for common text document sizes ranging from 1K to 1M characters, while $C_t$ is much larger than $C$ with the inclusion of tombstones (see Section 4.2). In practice, the following *inequality* commonly holds: $C_t \gg C \gg c$. These practical differences are often more significant than the theoretic differences in real world co-editing applications.

*5.3.3 Common Flaws in Characterizing OT and CRDT Theoretic Complexity.* To reveal the roots of CRDT false claims in complexity, we point out two common flaws in CRDT articles in characterizing OT and CRDT time and space complexities:

1. One flaw was to replace the variable $c$ (for OT) – the number of concurrent operations in the history buffer, with another variable of very different nature $H$ – the total number of operations executed in a co-editing session, which are accumulated in the history buffer and may grow large for long lasting sessions (more discussions on this in Section 5.3.4).

2. Another related flaw was to use the same $H$ to replace $C/C_t$ (for CRDT) – the number of objects in the internal object sequence. The trick here is that $C/C_t$ includes the objects for representing characters in the initial document, but $H$ does not capture the size of initial document contents at all. In fact, nearly all CRDT articles ignored the existence and impact of initial document contents in calculating the size of the internal object sequence and analyzing CRDT complexity.

With above variable replacements, the time complexity $O(c^2)$ or $O(c)$ for OT solutions were distorted into $O(H^2)$ or $O(H)$; and the time complexity $O(C_t^3)$, $O(C^2)/O(C_t^2)$, or $O(C)/O(C_t)$, for CRDT solutions became $O(H^3)$, $O(H^2)$, or $O(H)$, which, combined with additional twists, were the common root of flaws in CRDT theoretic complexity superiority claims [4,5,6,9,43,44,45,48,84,85,86].

*5.3.4 Issues Related to Non-Real-Time Co-Editing.* Some CRDT advocates[17] insist on that OT has to keep user-generated operations in the history buffer all the time, by citing scenarios in which co-editing users may be disconnected or offline during a session (in which concurrency-based garbage collection does not work), and argue CRDT may have an edge over OT under such circumstances. This argument is unjustified and false too.

First, supporting disconnected or offline co-editing is undoubtedly useful, but such co-editing scenarios belong to the class of *non-real-time* co-editing and should not be confused with *real-time* co-editing as there are important differences between these two classes of co-editing, such as different time and space constraints, different interaction and communication patterns, etc. [16,53,66]. Second, OT does not need to keep user-generated operations all the time under either real-time or non-real-time co-editing sessions. This is because OT has two special capabilities applicable to both real-time and non-real-time co-editing: (1) operation *compression* [53,54], which can merge multiple adjacent or overlapping inserts or deletes into a single operation or even eliminate them completely; (2) operation *composition* [83], which can serialize a stream of effective operations (that cannot be compressed) by their positional orders to make a single compound operation and highly efficient transformations can be applied between those compound operations. These operation compression and composition schemes are effective under both real-time and non-real-time co-editing and independent of concurrency-based garbage collection schemes. Finally, garbage collection based on concurrency is indeed ineffective under non-real-time co-editing, but such ineffectiveness is true for both operation garbage collection (in OT) and tombstone garbage collection (in CRDT).

A thorough treatment of issues in non-real-time co-editing, operation compression and composition is beyond the scope of this article. We refer readers to publications [16,52,53,54] and OTFAQ [66] (especially Q&A: "*what is the basic idea of OT for operation compression*", "*what issues to consider in evaluating OT system time efficiency*", and "*what are the space efficiency issues of an OT system*"), for detailed discussions on issues in designing and implementing OT-based non-real-time and hybrid real-time and non-real-time co-editors.

Apart from scattered claims about CRDT potential benefits under offline co-editing, to our best knowledge, there has been no published CRDT work specially addressing non-real-time co-editing issues or any CRDT-based working co-editor for supporting non-real-time co-editing [78].

## 5.4  Myths and Misconceptions in Simplicity Arguments of CRDT over OT
In conjunction with the notion that OT is complex, CRDT is often claimed to be simple. In this section, we tackle various versions and arguments about CRDT simplicity, which were scattered in published literature (e.g. [4,9,42,50,51,86]) and discussions among developers.

*5.4.1 CRDT Works without OT.* One version of the CRDT simplicity argument can be sketched as follows: CRDT works without OT, thus avoids complex issues associated with OT, hence CRDT is simple. The fallacies of this argument are: the simplicity of any approach cannot be logically argued on the basis of being different from or working without following another approach. The relevant questions that should be asked are: what special issues one approach brings in, whether those particular issues are easy to solve and have been solved, and whether solutions (if exist) to such issues are actually simple. In previous sections, we have provided ample evidences that show: CRDT has its own challenging issues and many of them remain unsolved (see Section 4); and CRDT solutions are not simple but more complex than OT solutions, as shown in Table 4.

---
[17] Examples can be found in discussions on OT and CRDT at https://news.ycombinator.com/item?id=18191867.

*5.4.2 The Myth of CRDT Native Commutativity by Design.* Another version of the CRDT simplicity is argued along the line how OT and CRDT achieve commutativity of concurrent operations in co-editors. CRDT was broadly defined as a data type or solution that makes "*concurrent operations commute with one another*" [42,50]; but the commutativity concept was not new at all as OT had been known for its capability of making concurrent editing operations commutative on replicated documents (see Section 2) long before CRDT came into being. To differentiate CRDT from OT, CRDT advocators formulated some arguments that can be sketched as follows: unlike OT which achieves the commutativity *after the fact* (by transformation), CRDT makes concurrent operations *natively* commutative and hence avoids inconsistency/conflict issues *by design*; avoiding inconsistency/conflicts or being *conflict-free* by design (CRDT) is more elegant and simpler than resolving them after the fact (OT) [50,51] (also see footnote 4).

Unfortunately, the CRDT native commutativity is a big myth ("*the emperor's new clothes*") in co-editing. The fact is: none of those CRDT components, including internal object structures and sequences (with/without tombstones), immutable identifier-based operations, and object sequence search and manipulation schemes, is native to any editors (see Section 2 and [78]). Despite its use of immutable identifier-based operations, CRDT cannot avoid the position-shifting issue (or be conflict-free) in text co-editing, but has to solve the same inconsistency/conflict issues and make position-based concurrent operations commutative by transformation (i.e. *after the fact*) as well, albeit *indirectly* (see Section 2), with much higher complexity than OT (see Section 4). The CRDT elegancy argument is contradicted by the facts that CRDT has inherently complex issues (e.g. $C/C_t$-based time and space complexity and tombstone overhead); representative CRDTs are intricate (e.g. the core *IntegrateIns* algorithm in WOOT, with $O(C_t^3)$ time complexity), and error-prone (e.g., flaws in the core identifier and object manipulation schemes in Logoot).

*5.4.3 Myths in OT/CRDT Implementation Complexity and Simplicity.* Still another notion of CRDT simplicity is related to implementation. OT was criticized by some for its implementation complexity while CRDT was argued to be simple to implement. Below, we examine what grounds (if any) on which those criticisms and arguments were based, and share our experiences in implementing OT and OT-based co-editors.

In over two decades, we have designed and implemented a range of OT solutions, including generic control algorithms GOT [58,60], GOTO [59,61], NICE [53], COT [71,72], TIBOT [88,89], and POT [88,89], and application-specific transformation functions for a variety of applications [2,3,12,60,66,70,74,75,76]. Moreover, we have applied those OT solutions to build over a dozen of publically demonstrable co-editing systems, including the Web-based pain-text co-editor REDUCE [60,62], desktop productivity tools like CoWord and CoPowerPoint [65,70,87], digital design tools like CoMaya [1,2], heterogeneous co-editing systems like CoVim+CoEmacs [10], and a range of Web-based rich-text co-editing products Codox Apps[18] [78].

From first-hand experiences, we have learnt that the bulk of co-editor implementation challenges lie mostly in applying OT to solving various co-editing issues and building co-editing systems, rather than in implementing OT algorithms themselves, which are at the core but only part of a co-editing system. While applying OT to building co-editors is non-trivial and could be time-consuming to people without prior knowledge of OT and co-editing issues, it should become significantly easier and quicker if one has done the work properly at least once and understood what had been done.

---

[18] https://www.codox.io

Nevertheless, OT was reported to be hard to implement in a well-known quote by a former Google Wave engineer:

"*Unfortunately, implementing OT sucks. [...] The algorithms are really hard and time consuming to implement correctly. [...] Wave took 2 years to write and if we rewrote it today, it would take almost as long to write a second time.*" [19]

This quote was often cited (and made available in the OT Wikipedia page[20]) by some to advocate the notion that OT is complex to implement and indirectly argue for CRDT simplicity, by following the logic − what is hard for OT will be easy for CRDT as CRDT works without OT.

However, what is less known and ignored is that the same engineer later amended the previous comments with the following:

"*For what it's worth, I no longer believe that wave would take 2 years to implement now - mostly because of advances in web frameworks and web browsers. When wave was written we didn't have websockets, IE9 was quite a new browser. We've come a really long way in the last few years.*" [21]

The above amended statements revealed that the major challenges of Google Wave were due to the Web frameworks, browsers, and communication utilities, etc. which were available at the time and chosen to build the OT-based Google Wave, rather than merely implementing the core Google Wave OT algorithms [39,83]. This reflection concurred with our experiences in implementing OT solutions and building OT-based co-editing systems, and with the fact that OT has been widely implemented and used in the vast majority of today's working co-editors.

On the other hand, the claim that CRDT is simple to implement was never substantiated by evidence, but only contradicted by the fact that CRDT is rarely implemented and used in today's co-editors. For detailed examination of OT and CRDT differences in implementation and building co-editors, the reader is referred to [78].

## 6  CONCLUSIONS

In this work, we have conducted comprehensive and critical reviews of representative OT and CRDT solutions for consistency maintenance in real-time co-editing, and made important research discoveries, which contribute to the advancement of the state-of-the-art knowledge on collaboration-enabling technology. This paper is the second in a series of three papers reporting our discoveries from this work. We summarize the main discoveries reported in this paper below, together with a brief description of the results reported in prior and follow-up papers as well.

In the first paper [78], we have presented a general transformation framework for examining and comparing OT and CRDT, among a variety of consistency maintenance solutions in co-editing, and revealed previously hidden but critical facts about CRDT: CRDT is like OT in following the same general transformation approach to consistency maintenance, albeit *indirectly* (in contrast to the *direct* transformation approach by OT); CRDT is the same as OT in making user-generated operations commutative *after the fact*; and CRDT operations are not natively commutative to editors, but require additional conversions between CRDT internal operations and external editing operations. Revealing these facts helps demystify what CRDT really *is* and *is not* to co-editing, and provide much-needed clarity for examining and understanding the real differences between OT and CRDT for co-editors − their radically different ways of realizing the same general transformation.

In the current paper, we have explored what really differentiates OT and CRDT in correctness and complexity. One key insight from this investigation is: OT is *concurrency-centric* in the sense it treats generic concurrency issues among operations as its first priority at the core control algorithms, and isolates the handling of application-specific data and operation modelling issues in transformation functions; whereas CRDT is *content-centric* in the sense that it takes the first priority to manipulate internal application-related contents, including object sequences and schemes for searching and applying identifier-based operations in the object sequence, but mixes the handling of concurrency issues within object search and manipulation schemes. This *concurrency-centric* vs *content-centric* difference, together with the *direct* vs *indirect* difference in transformation, are fundamental and have profound implications to OT and CRDT solutions.

---

[19] https://www.championtutor.com:7004/ (Nov 6, 2011)

[20] https://en.wikipedia.org/wiki/Operational_transformation

[21] https://news.ycombinator.com/item?id=12311984 (Aug 18, 2016). A reference to this link was later added to the OT Wikipedia page after our work reported in https://arxiv.org/abs/1810.02137.

The first significant implication is found in the different design and correctness issues in OT and CRDT solutions. Key OT design issues include designing core control algorithms to deal with generic concurrency issues, and designing separate transformation functions to handle application-specific issues. OT-special technical challenges and puzzles, such as ensuring context-based conditions (e.g. the *dOPT* puzzle was a case of violating the context-equivalence condition), and avoiding or preserving CP2 (e.g. the *FT* puzzle was a case of violating the CP2 property in plain-text co-editing), were rooted in and solved under the concurrency-centric approach. Past OT research has established context-based conditions and transformation properties as the theoretic foundation for verifying and guiding the design of OT solutions. The correctness of key OT components in major OT solutions, including generic control algorithms and transformation functions for a range of operation and data models (e.g. string-wise plain-text editing and beyond), has been established on top of this theoretic foundation.

In contrast, key CRDT design issues include designing CRDT-special data structures for representing the character sequence of the external document in an internal object sequence, immutable identifier-based operations, searching and executing identifier-based operations in the object sequence, and conversions between internal identifier-based operations and external position-based operations, which collectively deal with both application-specific and concurrency issues in co-editing. This approach has induced CRDT-specific challenges and puzzles, such as $C/C_t$-*based complexity, tombstone overhead, complexity with variable identifiers*, etc. In this work, we have revealed various correctness problems with Logoot (a representative non-tombstone-based CRDT), including *inconsistent-position-integer-ordering, infinite loop flaws, position-order-violation puzzles,* and *concurrent-insert-interleaving abnormalities*. It remains an open challenge to resolve those issues under the CRDT approach to co-editing. The correctness of key CRDT components (e.g. object sequences, operation identifiers, and object sequence searching and manipulation schemes) in various CRDT solutions remains to be verified, using well-defined criteria, which are yet to be established as well.

The second significant implication is found in the different time and space complexities among OT and CRDT solutions. OT complexity is determined by a variable $c$ (for *concurrency*) – the number of concurrent operations involved in transforming an operation; CRDT complexity is dominated by a variable $C$ (for *Contents*) or $C_t$ (for *Content with tombstones*) – the number of objects maintained in the internal object sequence. In terms of theoretic complexity (see details in Table 4), CRDT has time complexities $O(C)$ or $O(C_t)$ for local operation processing, whereas OT has $O(1)$ complexity. In terms of time complexity for remote processing and space complexity, both CRDT and OT have achieved linear complexities under respective variables $c$ (for OT) and $C/C_t$ (for CRDTs), which are seemingly similar in theory, but significantly different in practice: $c$ is often bounded by a small value, e.g. $0 \leq c \leq 10$, for a real-time session with a few (e.g. less than 5) users; $C$ is orders of magnitude larger than $c$, e.g. $10^3 \leq C \leq 10^6$, for common plain text document sizes ranging from 1K to 1M characters; and $C_t$ could be much larger than $C$ due to the inclusion of tombstones. In real-time text co-editing, the following inequality commonly holds: $\boldsymbol{C_t \gg C \gg c}$. It remains an open challenge to devise $\boldsymbol{C_t/C}$-based CRDT solutions that could match $\boldsymbol{c}$-based OT solutions in time and space complexity and in practical performance.

The third implication is in the generality and extendibility of OT and CRDT solutions for co-editors. OT solutions separate generic concurrency issues from application-specific data and operation issues, with the core control algorithms being generally applicable to different application domains beyond text editing. Extensions of existing OT solutions can be and have been achieved by designing new transformation functions for new applications, without reinventing core control algorithms. In contrast, CRDT solutions mix concurrency issues with application-specific data and operation issues, with key CRDT components being intricately related to each other and coupled with application-specific object sequences and operations. The vast majority of published CRDT solutions for co-editing have been confined to plain-text editing.

In the follow-up (the third) paper of this series [78], we report our discoveries on the differences between OT and CRDT in building co-editing systems and real world applications. In particular, we investigate the role of building co-editing systems in shaping OT and CRDT research and solutions. We review the evolution of co-editors from research vehicles to real world applications, and discuss representative OT-based co-editors and alternative design approaches in industry products and open source projects. Moreover, we evaluate CRDT-based co-editors in relation to published CRDT solutions, and re-confirm that CRDT operations and object sequences are not native to any editors. Last but not least, we examine technical factors related to "peer-to-peer" co-editing, such as the use of central servers, causally-ordered communication, and vector/scalar timestamping. Our study revealed all these factors are orthogonal to OT and CRDT, and the notion that CRDT is especially suitable for supporting "peer-to-peer" co-editing is a fallacy.

In summary, we have critically reviewed and compared representative OT and CRDT solutions, with respect to their basic approaches to consistency maintenance, correctness, time and space complexity, implementation and application in building co-editors, and suitability for peer-to-peer co-editing. The facts and evidences revealed from this work disprove CRDT superiority claims over OT on all accounts. These results help explain the underlying reasons behind the choices between OT and CRDT in real world co-editors and industrial products.

Numerous alternative consistency maintenance solutions have been explored in past co-editing research, and a wealth of experiences and lessons have been accumulated from those explorations. The time is ripe to review them critically and reflect on: what each alternative really is and has achieved so far, whether it has been validated by or is relevant to real world applications, and whither it is heading. This critical review work on OT and CRDT represents one attempt in this direction. For any alternative to be a viable solution in co-editing, in our view, it should offer capabilities that are genuinely superior to existing state-of-the-art solutions, and demonstrate its relevance in supporting real world co-editors. It is the real world application that provides ultimate validation to alternative solutions and to co-editing research in general.

We hope discoveries from this work will help clear up common myths and misconceptions surrounding OT and CRDT, inspire fruitful explorations of novel collaboration techniques, and accelerate progress in co-editing technology innovation and real world application.

## ACKNOWLEGMENTS

This research is partially supported by Academic Research Fund Tier 2 (MOE2015-T2-1-087) Grant from Ministry of Education, Singapore. The authors are grateful to the anonymous expert reviewers for their insightful and constructive comments and suggestions, which have helped improve the presentation of this article.